\documentclass[trackchanges,times,tighten,twocolumn]{aastex701}

\usepackage{booktabs}

\begin{document}

\title{Worlds Next Door. V. A Candidate Solar System Scale Super-Jupiter in the 61 Cygni Binary System}

\correspondingauthor{Aniket Sanghi}

\author[0000-0002-1838-4757]{Aniket Sanghi}
\altaffiliation{NSF Graduate Research Fellow}
\affiliation{Cahill Center for Astronomy and Astrophysics, California Institute of Technology, 1200 E. California Boulevard, MC 249-17, Pasadena, CA 91125, USA}
\email[show]{asanghi@caltech.edu}

\author[0000-0002-9644-8330]{Billy Quarles}
\affiliation{Department of Physics and Astronomy, East Texas A\&M University Commerce, TX 75428, USA}
\email{billylquarles@gmail.com}

\author[0000-0002-9082-6337]{Andrea S.J. Lin}
\affiliation{Cahill Center for Astronomy and Astrophysics, California Institute of Technology, 1200 E. California Boulevard, MC 249-17, Pasadena, CA 91125, USA}
\email{asjlin@caltech.edu}

\author[0000-0003-1247-9349]{Cheyanne Shariat}
\affiliation{Cahill Center for Astronomy and Astrophysics, California Institute of Technology, 1200 E. California Boulevard, MC 249-17, Pasadena, CA 91125, USA}
\email{cshariat@caltech.edu}

\author[0000-0003-0626-1749]{Pierre Kervella}
\affiliation{LIRA, Observatoire de Paris, Universit\'e PSL, Sorbonne Universit\'e, Universit\'e Paris Cit\'e, CY Cergy Paris Universit\'e, CNRS, 5 place Jules Janssen, 92195 Meudon, France}
\affiliation{French-Chilean Laboratory for Astronomy, IRL 3386, CNRS and U. de Chile, Casilla 36-D, Santiago, Chile}
\email{pierre.kervella@obspm.fr}

\author[0000-0001-5864-9599]{James Mang}
\altaffiliation{NSF Graduate Research Fellow}
\affiliation{Department of Astronomy, University of Texas at Austin, Austin, TX 78712, USA}
\email{j_mang@utexas.edu}

\author[0000-0002-8895-4735]{Dimitri Mawet}
\affiliation{Cahill Center for Astronomy and Astrophysics, California Institute of Technology, 1200 E. California Boulevard, MC 249-17, Pasadena, CA 91125, USA}
\affiliation{Jet Propulsion Laboratory, California Institute of Technology, Pasadena, CA 91109, USA}
\email{dmawet@astro.caltech.edu}

\author[0000-0002-0078-5288]{Mark R. Giovinazzi}
\affiliation{Department of Physics and Astronomy, Amherst College, 25 East Drive, Amherst, MA 01002, USA}
\email{mgiovinazzi@amherst.edu}

\author[0000-0002-5463-9980]{Arvind F. Gupta}
\affiliation{U.S. National Science Foundation National Optical-Infrared Astronomy Research Laboratory, 950 N. Cherry Avenue, Tucson, AZ 85719, USA}
\email{arvind.gupta@noirlab.edu}

\author[0000-0002-5627-5471]{Charles Beichman}
\affiliation{NASA Exoplanet Science Institute, Caltech-IPAC, Pasadena, CA 91125, USA}
\affiliation{Jet Propulsion Laboratory, California Institute of Technology, Pasadena, CA 91109, USA}
\email{chas@ipac.caltech.edu}

\author[0000-0003-4205-4800]{Bertrand Mennesson}
\affiliation{Jet Propulsion Laboratory, California Institute of Technology, Pasadena, CA 91109, USA}
\email{bertrand.mennesson@jpl.nasa.gov}

\author[0000-0002-2918-8479]{Mathilde M\^alin}
\affiliation{Space Telescope Science Institute, 3700 San Martin Drive, Baltimore, MD 21218, USA}
\affiliation{Department of Physics \& Astronomy, Johns Hopkins University, 3400 N. Charles Street, Baltimore, MD 21218, USA}
\email{mmalin@stsci.edu}

\author[0000-0002-6618-1137]{Jerry W. Xuan}
\altaffiliation{51 Pegasi b Fellow}
\affiliation{Department of Physics, University of California, Santa Barbara, Santa Barbara, CA 93106, USA}
\affiliation{Department of Earth, Planetary, and Space Sciences, University of California, Los Angeles, CA 90095, USA}
\email{jerryxuan@g.ucla.edu}

\author[0000-0001-7031-8039]{Liton Majumdar}
\affiliation{Exoplanets and Planetary Formation Group, School of Earth and Planetary Sciences, National Institute of Science Education and Research, Jatni 752050, Odisha, India}
\affiliation{Homi Bhabha National Institute, Training School Complex, Anushaktinagar, Mumbai 400094, India}
\email{dr.liton.majumdar@gmail.com}

\author[0000-0001-6513-1659]{Jack J. Lissauer}
\affiliation{Space Science \& Astrobiology Division, Building 245, NASA Ames Research Center, Moffett Field, CA 94035, USA}
\email{jack.lissauer@nasa.gov}

\author[0000-0001-9064-5598]{Mark Wyatt}
\affiliation{Institute of Astronomy, University of Cambridge, Madingley Road, Cambridge CB3 0HA, UK}
\email{wyatt@ast.cam.ac.uk}

\author[0000-0001-6396-8439]{William O. Balmer}
\affiliation{Department of Physics \& Astronomy, Johns Hopkins University, 3400 N. Charles Street, Baltimore, MD 21218, USA}
\affiliation{Space Telescope Science Institute, 3700 San Martin Drive, Baltimore, MD 21218, USA}
\email{wbalmer1@jhu.edu}

\author[0000-0002-9408-8925]{Eduardo Bendek}
\affiliation{NASA Ames Research Center, Moffett Field, CA 94035, USA}
\email{eduardobendek@gmail.com}

\author[0000-0002-4309-6343]{Kevin Wagner}
\affiliation{Department of Astronomy and Steward Observatory, University of Arizona, USA}
\email{kevinwagner@arizona.edu}

\begin{abstract}
We present new constraints on a candidate third companion in the 61 Cygni binary system---one of the closest neighbors to our solar system (3.5~pc), the first stars to have a measured parallax, and two of the most favorable targets for habitable zone exo-Earth searches with future direct imaging missions. \citet{kervella_stellar_2022} first identified the candidate based on a 4.4$\sigma$ significance tangential velocity anomaly in the system determined using Hipparcos and Gaia DR3 astrometry of the A and B components. We combine this astrometric evidence with constraints from Gaia DR3 {\tt RUWE} measurements, MMT/Clio $Lp$ adaptive optics imaging, nearly four decades of radial velocity (RV) measurements (including $\sim$2~years of new NEID RVs presented here), and $N$-body dynamical stability calculations of both stars in a Monte Carlo simulation framework to estimate the candidate companion's mass and semi-major axis. Our analysis reveals that all of the available data are compatible with an $\approx(8 \pm 3)\;M_{\rm Jup}$ super-Jupiter in an $\approx(8 \pm 3)$~au orbit, with a broad range of allowed eccentricities and mutual inclinations. It cannot be determined which component of the binary the candidate orbits. Using the Sonora Flame Skimmer evolutionary models, given the system's mature age of $(6.0 \pm 1.0)$~Gyr, we estimate a companion effective temperature of $(282 \pm 55)$~K and radius of $(1.02 \pm 0.02)\;R_{\rm Jup}$. Finally, we evaluate the stability of terrestrial planets in the habitable zone (HZ) of 61~Cygni~AB in the presence of the giant planet candidate using $N$-body simulations. A companion semi-major axis $\lesssim$5~au, irrespective of the planet-planet mutual inclination, and between $\approx$5~au and 10~au, for misaligned orbits ($40^\circ \lesssim i_{\rm mut} \lesssim 140^\circ$), severely disrupts the HZ. Our results highlight the need to confirm the planet candidate and characterize its orbit to inform future exo-Earth searches with missions such as SHERA, HWO, and LIFE.
\end{abstract}


\section{Introduction} 

Giant planets play a critical role in the formation, evolution, and habitability of terrestrial exoplanets \citep[e.g.][]{raymond_search_2006}. At early times, they can restrict the inward migration of protoplanetary cores that would deplete accretion material in the habitable zone \citep[HZ;][]{izidoro_terrestrial_2014, izidoro_gas_2015}. They may facilitate the delivery of life-supporting volatiles to terrestrial planets by influencing the rate of cometary and/or minor body impacts \citep{horner_jupiter_2008, horner_jupiter_2009, raymond_origin_2017, venturini_setting_2020, kane_planetesimal_2025}. A giant planet near the HZ can also destabilize potential terrestrial planets in the region over long timescales through dynamical perturbations \citep[e.g.,][]{noble_orbital_2002, kane_dynamical_2024, beichman_worlds_2025}. Thus, it is necessary to detect and characterize, or place limits on the presence of, any massive planet(s) in nearby systems most likely to be targeted in the search for habitable exoplanets.

At a distance of 3.5~pc, the 61~Cygni system (hereafter 61~Cyg; Table~\ref{tab:prop}) is one of the nearest neighbors to our solar system. The main sequence components 61~Cyg~A (K5V, $\approx0.68\;M_\odot$) and 61~Cyg~B (K7V, $\approx0.63\;M_\odot$) together form an eccentric visual binary in a long-period orbit \citep[$e \approx 0.44$, $a \approx 87$~au, $P \approx 707$~year;][]{giovinazzi_neid_2025}. 61 Cyg's age is estimated to be $6.0 \pm 1.0$~Gyr \citep{kervella_radii_2008}, similar to our solar system's age of $\sim$4.6~Gyr. The system exhibits the $6^{\rm th}$ largest proper motion, first estimated by \citet{Piazzi1806} \citep[][]{ fodera_serio_giuseppe_1990}, among all stars cataloged by Gaia \citep[$\approx 5\farcs3$/yr;][]{gaia_collaboration_gaia_2023}. This proper motion measurement led to one of the most significant achievements in astrophysics, the first determination of a stellar parallax \citep{bessel_ii_1838}. Nearly two centuries later, 61~Cyg~AB continue to hold importance in astrophysics. They are two of the closest and most favorable targets \citep[Tier~A;][]{mamajek_nasa_2024} to search for, directly image \citep[amenable to near 100\% search completeness;][]{gaudi_habitable_2020}, and detect biosignatures \citep{arney_k_2019} in HZ Earth-analogs with future missions such as the Searching for Habitable Exoplanets with Relative Astrometry \citep[SHERA;][]{christiansen_searching_2026} mission concept, the Habitable Worlds Observatory \citep[HWO;][]{feinberg_habitable_2026}, and the Large Interferometer for Exoplanets \citep[LIFE;][]{quanz_large_2022} mission. 

\begin{figure*}
    \centering
    \includegraphics[width=\linewidth]{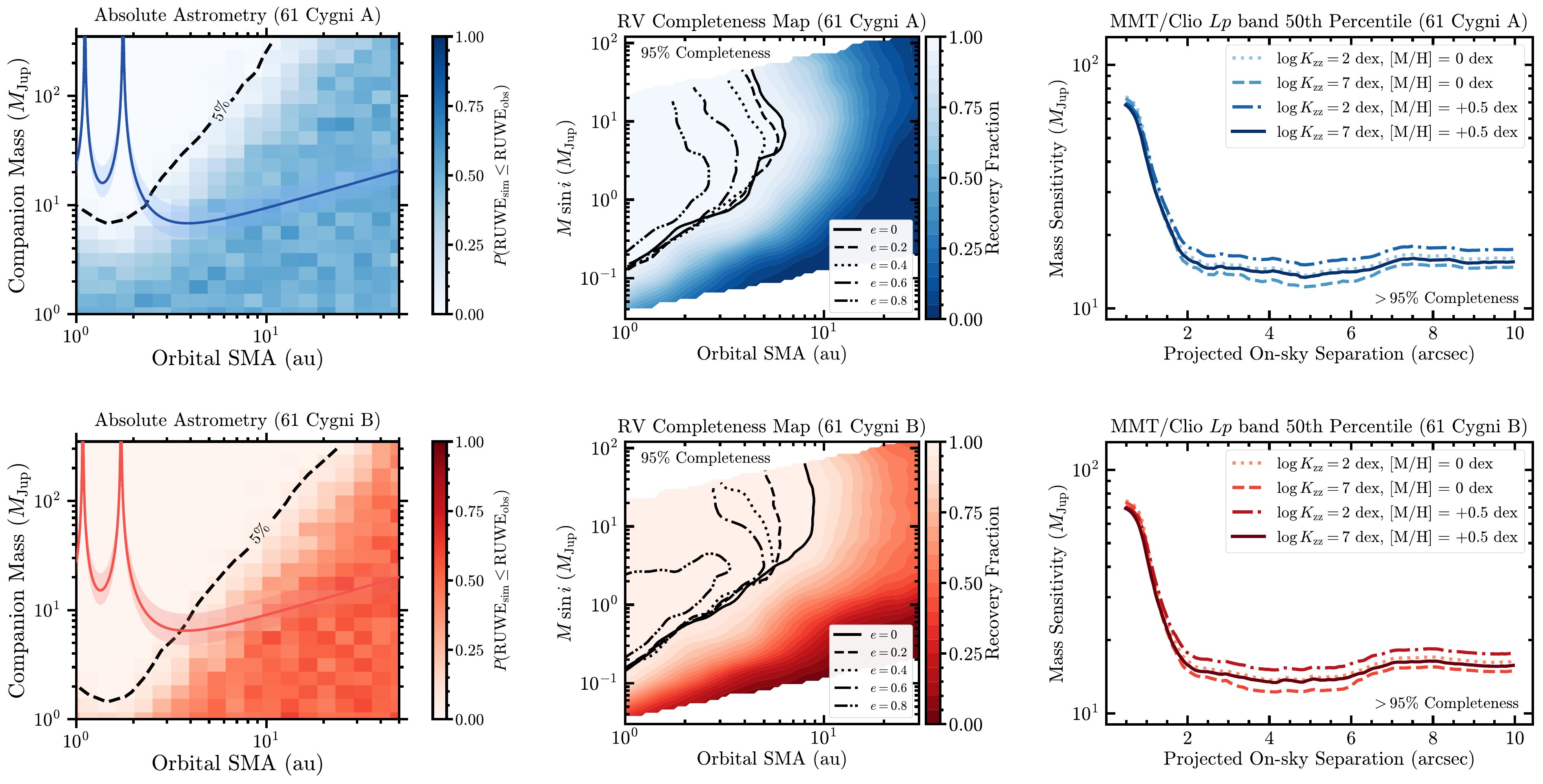}
    \caption{Datasets used in this work to constrain the presence of companions in the 61~Cyg~AB system. \emph{Left:} Companion mass vs.\ orbital semi-major axis (SMA, solid line) that explains the orbital velocity anomaly ($\Delta v_{\rm orb} = 88.5\pm19.8$ m/s) of 61~Cyg~AB \citep{kervella_stellar_2022}. It is not known which star the companion orbits. The background grid shows the fraction of simulated orbits in each cell with a {\tt RUWE} less than or equal to the observed {\tt RUWE} for each star. The 5\% probability contour (adopted here) is shown as a dashed curve. \emph{Center:} RV detection completeness map for $e = 0$ planets (most optimistic case) orbiting 61~Cyg~A and B based on archival RV observations from Lick/Hamilton, HIRES, APF, MINERVA, and NEID. The 95\% completeness contours (adopted for the analysis in this paper) for various orbital eccentricity values are overlaid as black lines. \emph{Right:} Companion mass sensitivity ($>$95\% completeness) as a function of projected on-sky separation based on direct imaging observations with MMT/Clio in the $Lp$ band \citep[][]{heinze_constraints_2010} for various metallicities and vertical mixing strengths.}
    \label{fig:data}
\end{figure*}

61~Cyg~AB have been frequently targeted in the search for exoplanets \citep[][]{strand_61_1943, strand_orbital_1957, deich_invisible_1977, heintz_reexamination_1978, walker_search_1995, cumming_lick_1999, cumming_keck_2008, heinze_constraints_2010, howard_limits_2016, butler_lces_2017, rosenthal_california_2021, hirsch_understanding_2021}, however, there are no confirmed detections to date. Recently, \citet{kervella_stellar_2019, kervella_stellar_2022} presented evidence for the presence of a third companion in the system based on Hipparcos and Gaia astrometry of the two stars. They find a companion mass $\approx 10\;M_{\rm Jup}$ and a semi-major axis between 10~au and 20~au (around either star) based on a rough analysis incorporating radial velocity (RV) data from \citet{howard_limits_2016}, imaging data from \citet{heinze_constraints_2010}, and the stability domain from \citet{musielak_stability_2005}. Motivated by the above work, we aim to address two key questions in this paper with the addition of new RV data and detailed numerical stability calculations: 
\begin{enumerate}
    \item What are the physical and orbital properties of the candidate companion?
    \item How does the candidate companion dynamically influence a potential terrestrial planet in 61~Cyg~AB's HZ?
\end{enumerate}

The paper is thus organized as follows. Section~\ref{sec:data} summarizes the astrometric evidence for a companion in the 61~Cyg system and presents archival imaging observations as well as improved sensitivity estimates to planets in the system with new NEID RV observations. Section~\ref{sec:constraints} combines absolute astrometry, RV, imaging, and stability calculations to place new constraints on the orbital and physical properties of the companion in the system. Section~\ref{sec:stable-earths} investigates the stability of HZ terrestrial planets in the presence of the companion inferred from astrometry. Section~\ref{sec:concl} lists our conclusions. Finally, Appendix~\ref{app:rv} presents the fits to the complete RV datasets of 61~Cyg~AB.

\begin{deluxetable}{lccc}
\tabletypesize{\footnotesize}
\tablecaption{\label{tab:prop}Stellar Properties of 61~Cyg~AB}
\tablehead{\colhead{Property} & \colhead{61 Cyg A} & \colhead{61 Cyg B} & \colhead{Ref.}}
\startdata
$\alpha_{2016.0}$ ($^\circ$) & 316.7485 $\pm$ 0.0388 & 316.7537 $\pm$ 0.0190 & 1\\
$\delta_{2016.0}$ ($^\circ$) & +38.7639 $\pm$ 0.0485 & +38.7561 $\pm$ 0.0232 & 1\\
$\mu_{\alpha^*}$\tablenotemark{a} (mas\,$\mathrm{yr^{-1}}$) & 4164.209 $\pm$ 0.055 & 4105.976 $\pm$ 0.026 & 1 \\
$\mu_{\delta}$ (mas\,$\mathrm{yr^{-1}}$) & 3249.614 $\pm$ 0.055 & 3155.942 $\pm$ 0.027 & 1 \\
$\varpi$ (mas) & $285.9949 \pm 0.0599$ & $286.0054 \pm 0.0289$ & 1\\
Distance (pc) & $3.49624^{+0.00076}_{-0.00062}$ & $3.49607^{+0.00034}_{-0.00037}$ & 2\\
SpT & K5V & K7V & 3\\
Mass ($M_\odot$) & $0.6771^{+0.0051}_{-0.0051}$ & $0.6289^{+0.0094}_{-0.0092}$ & 4 \\
Radius ($R_{\odot}$) & 0.667 $\pm$ 0.005 & 0.594 $\pm$ 0.008 & 5 \\
Age (Gyr) & $6.0 \pm 1.0$ & $6.0 \pm 1.0$ & 5 \\
$T_{\mathrm{eff}}$ (K) & 4398 $\pm$ 34 & 4174 $\pm$ 47 & 6 \\
$\mathrm{[Fe/H]}$ (dex) & $-$0.13 $\pm$ 0.03 & $-$0.21 $\pm$ 0.07 & 6 \\
$\log(R'_{HK})$ (dex) & --4.588 & --4.803 & 7\\
$P_\mathrm{rot}$ (d) & 35.54 $\pm$ 0.47 & 34.55 $\pm$ 0.5 & 7\\
$\mathrm{RUWE_{DR3}}$ & 1.204 & 0.961 & 1\\
\enddata
\tablenotetext{a}{Proper motion in R.A. includes a factor of $\cos \delta$.}
\tablerefs{(1)~\citet{gaia_collaboration_gaia_2023}; (2)~\citet{bailer-jones_estimating_2021}; (3)~\citet{keenan_perkins_1989}; (4)~\citet{giovinazzi_neid_2025}; (5)~\citet{kervella_radii_2008}; (6)~\citet{soubiran_gaia_2024}; (7)~\citet{olspert_estimating_2018}.
}
\end{deluxetable}

\section{Data}
\label{sec:data}

\subsection{Hipparcos and Gaia Absolute Astrometry}
\label{sec:astrometry}

The astrometric evidence for a companion in the 61~Cygni system using Hipparcos and Gaia absolute astrometry was first presented by \citet{kervella_stellar_2019, kervella_stellar_2022}. Here, we provide a brief summary of the argument presented in \citet{kervella_stellar_2022}. 

For a binary star system, the proper motion vector of the system barycenter can be estimated using two different methods: (1)~from the mass-weighted mean of the proper motion vectors of each component at the DR3 epoch (short-term proper motion) and; (2)~from the difference in barycenter position (computed as the mass-weighted mean of individual component positions) between the Hipparcos and Gaia DR3 epochs (long-term proper motion). \citet{kervella_stellar_2022} find a difference at the 3$\sigma$ level between the barycenter proper motion calculated using the above two approaches (see their Table~5). This proper motion anomaly likely indicates a bias in the short-term barycenter proper motion from the gravitational effect of a third companion orbiting one of the two stars (S-type configuration). A circumbinary (P-type) configuration is unlikely since it would correspond to a millenial scale orbit that cannot explain the shift in the short-term proper motion of the barycenter.

\citet{kervella_stellar_2022} additionally compute the orbital velocity vectors of each component as the difference between the Gaia proper motion (of a given star) and the long-term Hipparcos-Gaia barycentric proper motion. The tangential velocity vectors are expected to be colinear (offset by 180$^\circ$) with magnitudes inversely proportional to the component masses for a simple two-star system. In this scenario, the differential quantity, $\Delta v_{\rm orb} = v_B - \frac{M_A}{M_B}\cdot v_A = 0$, where $\Delta v_{\rm orb}$ is the orbital velocity anomaly vector, $v_A$ and $v_B$ are the orbital velocities of the two components, and $M_A$ and $M_B$ are the masses of the two components. For 61 Cygni, \citet{kervella_stellar_2022} find an orbital velocity anomaly $\Delta v_{\rm orb} = 88.5 \pm 19.8$~m/s, significant at the 4.4$\sigma$ level. The interpretation of the orbital velocity anomaly is similar to that of the proper motion anomaly for a single star \citep{brandt_hipparcos-gaia_2021, kervella_stellar_2022} and results in a mass-semimajor axis degeneracy for the potential companion responsible for the same (Figure~\ref{fig:data}). As the measurement is differential in nature, it is not possible to tell which star the companion orbits. We will use this as the starting point for the candidate companion's properties in \S\ref{sec:constraints}.

\subsection{Constraints from Gaia DR3 RUWE}
\label{sec:RUWE}

The Gaia DR3 renormalized unit weight error ({\tt RUWE}) quantifies the quality of the single-star astrometric fit. By construction, well-modeled single stars have ${\tt RUWE}\approx$~1 \citep{lindegren_gaia_2018}. {\tt RUWE} can be used as an independent astrometric constraint on the presence of faint orbiting companions. A close companion capable of producing the velocity anomaly measured by \citet{kervella_stellar_2022} should also move the photocenter during the Gaia DR3 observing window and thus increase the residuals of the single-source astrometric solution \citep{gaia_collaboration_gaia_2023,el-badry_generative_2024}. Gaia DR3 reports ${\tt RUWE}=1.204$ for 61~Cyg~A and ${\tt RUWE}=0.961$ for 61~Cyg~B \citep{gaia_collaboration_gaia_2023}. We use \texttt{gaiamock}\footnote{\url{https://github.com/kareemelbadry/gaiamock}} to forward-model the DR3 observations, treating A and B separately as possible host stars \citep{el-badry_generative_2024}. We evaluated a $21\times21$ grid spanning $0.6$--$50$ au and $1$--$300\,M_{\rm Jup}$, with 128 realizations per grid cell. Each realization assumes a dark, unresolved companion, isotropic orientations, uniform orbital angles and phase, and $e\sim\mathcal{U}(0,0.99)$. We consider the orbital motion of the 61~Cyg~AB binary during this period using the orbital constraints of \citet{giovinazzi_neid_2025}, although we find it to be negligible over Gaia DR3’s 3-year baseline. We then calculate $P({\tt RUWE}_{\rm sim}\leq{\tt RUWE}_{\rm obs}\mid m_p, a)$ at every grid point. A probability near 0.5 indicates that the companion produces no detectable astrometric perturbation, whereas a probability near zero means that nearly all simulated realizations have a higher RUWE than observed and are therefore inconsistent with the Gaia DR3 data.

The resulting probability map is shown in the left panel of Figure~\ref{fig:data}. 
For companions within $\sim2$~au, the probability of reproducing the observed RUWE is $<5\%$ for $\gtrsim10$--$20\,M_{\rm Jup}$ mass companions around 61~Cyg~A and $\gtrsim2$--$3\,M_{\rm Jup}$ around 61~Cyg~B. The two narrow, high-mass spikes near $1.1$~au and $1.7$~au are therefore strongly disfavored, whereas the lower-mass branch beyond several au remains compatible with the Gaia data. The tighter limits for companions around 61~Cyg~B arise mainly from its smaller DR3 astrometric uncertainties.

\subsection{Ground-based Adaptive Optics Imaging}
\label{sec:imaging}
61~Cyg~A and B were observed by \citet{heinze_constraints_2010} with the Clio Camera on the MMT Observatory \citep{freed_clio_2004} as part of an $Lp$- and $M$-band adaptive optics imaging survey of nearby Sun-like stars for exoplanets. \citet{heinze_constraints_2010} adopt a nominal 10$\sigma$ detection threshold (in apparent magnitudes; corresponding to a completeness of $97\%$) and report sensitivity as a function of the two-dimensional (2D) ($\Delta$R.A., $\Delta$Decl.) on-sky position of the companion. Using these 2D maps, they construct one-dimensional (1D) sensitivity curves for various percentiles in sensitivity as a function of separation. In our analysis, we use the 50th percentile curve, which corresponds to the median (97\% completeness) sensitivity across all position angles at a fixed separation. 

\begin{figure}
    \centering
    \includegraphics[width=\linewidth]{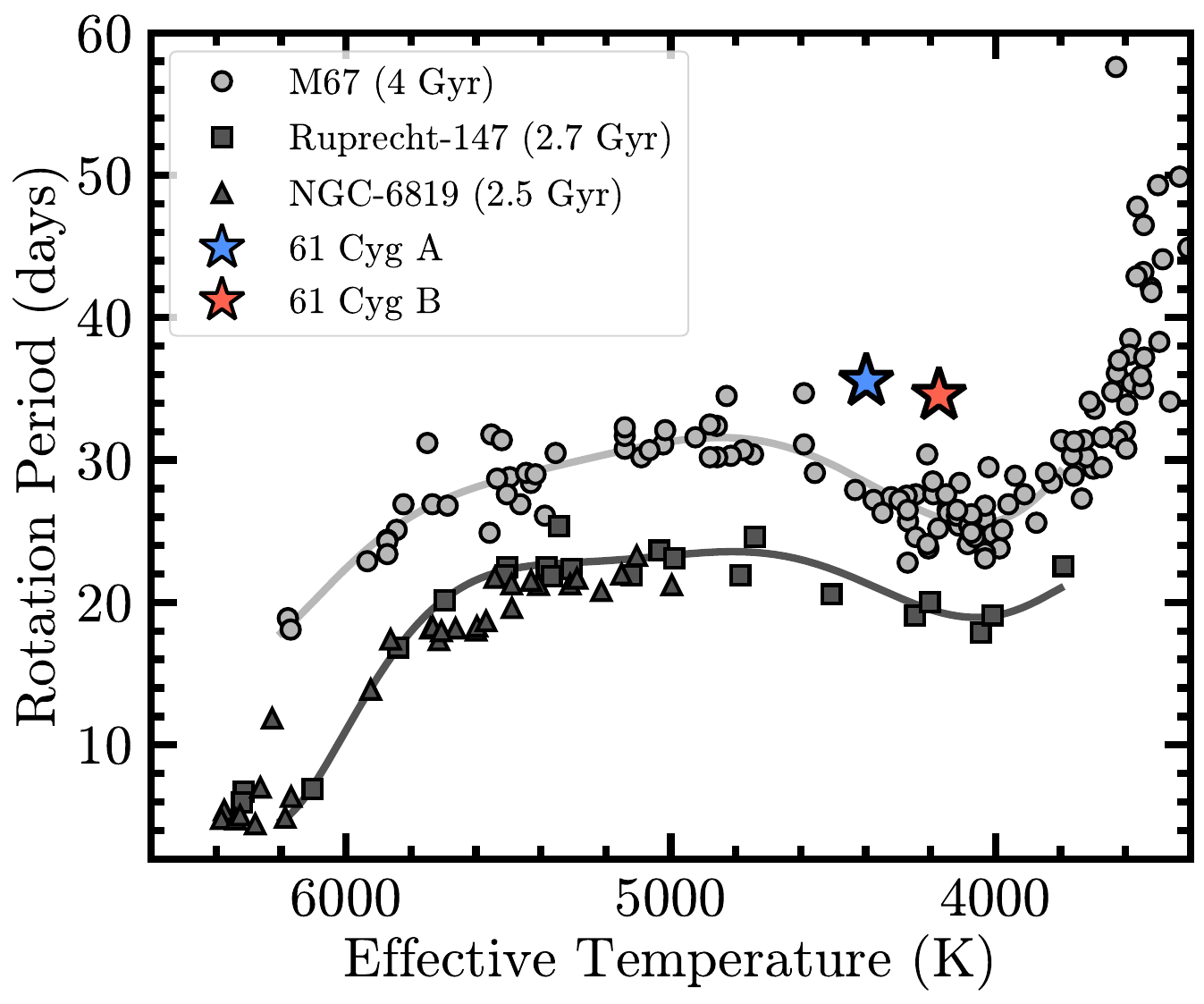}
    \caption{Effective temperature vs.~rotation period for 61~Cyg~A/B (Table~\ref{tab:prop}; reported uncertainties are smaller than the marker size) compared to data from three age-dated clusters compiled by \citet{bouma_empirical_2023}. Solid lines show polynomial fits to the data. Original sources of cluster data: M67 \citep{gruner_new_2023}, Ruprecht-147 \citep{curtis_when_2020}, and NGC-6819 \citep{meibom_spin-down_2015}.}
    \label{fig:gyro}
\end{figure}

\citet{heinze_constraints_2010} convert the sensitivity in apparent magnitudes to companion mass using the \citet{burrows_beyond_2003} evolutionary models for an age of 2~Gyr, the default assumed based on dynamical constraints on the mean age thin-disk stars in the solar neighborhood, for stars where they did not find specific age estimates. However, using evolutionary models that reproduce the interferometrically-determined radii of both stars in the binary, \citet{kervella_radii_2008} find an age of $6.0 \pm 1.0$~Gyr. This differs from previous age estimates derived from gyrochronology \citep[$\approx$3.5~Gyr;][]{mamajek_improved_2008} and chromospheric activity \citep[2.4--3.8~Gyr;][]{barnes_ages_2007}. In the past few years, new measurements of rotation periods for low-mass members of age-dated star clusters (primarily from Kepler and TESS) has led to refined calibration of gyrochronology relations \citep[e.g.,][]{bouma_empirical_2023}. Using 61~Cyg~A/B's effective temperature from \citet{soubiran_gaia_2024} and rotation period measurements from \citet{olspert_estimating_2018}, we compare the two stars to members of three age-dated clusters (2.5--4~Gyr, Figure~\ref{fig:gyro}) as compiled by \citet{bouma_empirical_2023}\footnote{Data for the M67 cluster from \citet{gruner_new_2023} was added in \texttt{gyro-interp} v0.4 \citep{bouma_gyrointerp_2023}.}. We do not attempt to derive a posterior distribution for the age with \texttt{gyro-interp} \citep{bouma_gyrointerp_2023} as the stars fall in the region of model extrapolation. Nevertheless, the position of both stars on the diagram does support an $\approx$solar-age ($\gtrsim4$~Gyr) for the system. High-cadence radial velocity observations (e.g., with VLT/ESPRESSO, Keck Planet Finder, HARPS; $\approx$45--60~second exposures) offer a pathway to detect solar-like oscillations in cool K dwarfs and derive an asteroseismic age for the two stars \citep[see example detections for $\alpha$~Cen~B, $\epsilon$~Ind~A, and $\sigma$~Draconis,][]{kjeldsen_solar-like_2005, campante_expanding_2024, lundkvist_low-amplitude_2024, hon_asteroseismology_2024}.

We adopt an age of 6~Gyr to use the Clio imaging non-detection to constrain possible masses and orbits of the proposed candidate companion. An older age reduces the detectability of lower-mass (colder) planets for a fixed magnitude sensitivity level and thus is the conservative choice. For this age, we convert the companion flux sensitivity (in apparent magnitudes) of the $Lp$-band observations (acquired on 2006 June~09 for 61~Cyg~A and 2006 June~10 for 61~Cyg~B) to mass limits using Sonora Flame Skimmer atmospheric and evolutionary models \citep{mang_sonora_2026} (described in \S\ref{sec:teff-rad}). We tested weak and strong vertical mixing scenarios in combination with solar and enhanced atmospheric metallicity scenarios for a C/O ratio $= 1\times$~solar, as relevant for objects in the $\gtrsim10\;M_{\rm Jup}$ mass range \citep[e.g.,][]{miles_observations_2020, xuan_are_2024, chachan_revising_2025}. Figure~\ref{fig:data} shows that mass sensitivity does not change significantly across these parameters\footnote{The mass sensitivity obtained with the $M$-band contrast curves depends strongly on the $\log K_{\rm zz}$ mixing parameter due to CO absorption at $\approx$4.6~$\mu$m. Thus, we do not use them in our analysis.}. We adopt the mass sensitivity curve calculated using the $+0.5$~dex metallicity, $\log K_{\rm zz} = 7$~dex, solar C/O ratio Sonora Flame Skimmer model when incorporating the imaging constraints in our analysis.

\subsection{Radial Velocities}
\label{sec:RV}

As previously mentioned, 61~Cyg~AB have long been popular targets for exoplanet searches, and they have been monitored by planet-hunting RV instruments since the advent of the technique in the late 1980s. \citet{howard_limits_2016} used nearly 30 years of accumulated RVs from the Lick/Hamilton \citep{Vogt1987_lick} and Keck/HIRES \citep{Vogt1994_hires} spectrographs to put constraints on the masses of potential planets in the system. They found that 61~Cyg A and B both showed approximately linear RV trends due to the orbit of the other star. Once these trends were removed, they were able to rule out planets orbiting either star (circumbinary planets were not considered) with $M_p \sin{i} \gtrsim 0.3\ M_J$ within $\sim$3~au and $\gtrsim 10\ M_J$ within $\sim$10~au through injection-recovery tests. These tests estimate detection completeness by injecting synthetic RV signals into the pre-existing dataset and attempting to recover them using the same algorithms used to search for potential real planets.

\begin{deluxetable}{lccccccccc}
\label{tab:rvs}
\tabletypesize{\footnotesize}
\centering
\tablecaption{Summary of RVs}
\tablehead{
~          & \multicolumn{4}{c}{61~Cyg~A}          & \multicolumn{4}{c}{61~Cyg~B}          & \\
\cmidrule(lr){2-5} \cmidrule(lr){6-9}
Instrument & $N_{RV}$ & Med. error   & Start & End & $N_{RV}$ & Med. error     & Start & End & Source \\
~          &          & (m\,s$^{-1}$) &       &     &          & (m\,s$^{-1}$) &       &     & }
\startdata
Hamilton                       & 109  & 8.31  & 1987-06-11  & 2006-12-03  & 70   & 8.15  & 1987-06-12  & 2007-11-22 & 1 \\
HIRES (post)\tablenotemark{a}  & 298  & 1.26   & 2004-08-21  & 2022-10-18  & 281  & 0.89   & 2004-08-21  & 2022-10-18 & 2 \\
APF                            & 345  & 2.62  & 2013-10-17  & 2019-03-15  & 259  & 2.63  & 2013-11-01  & 2019-03-15 & 1 \\
MINERVA \\
~~~~T1                         & 52   & 1.97  & 2017-09-17  & 2019-11-12  & \nodata & \nodata & \nodata & \nodata & 3 \\
~~~~T2                         & 33   & 4.52  & 2018-10-11  & 2019-11-12  & \nodata & \nodata & \nodata & \nodata & 3 \\
~~~~T3                         & 22   & 4.08  & 2017-10-05  & 2019-12-14  & \nodata & \nodata & \nodata & \nodata & 3 \\
~~~~T4                         & 51   & 2.08  & 2017-09-17  & 2019-10-21  & \nodata & \nodata & \nodata & \nodata & 3 \\
NEID (\texttt{SERVAL}) \\
~~~~Run 1                      & 86   & 0.18  & 2021-05-31  & 2022-06-07  & 40   & 0.13  & 2021-10-06  & 2022-06-10 & 4 \\
~~~~Run 2                      & 100  & 0.18  & 2022-12-01  & 2024-06-11  & 52   & 0.16  & 2022-11-12  & 2024-05-30 & 4 \\
~~~~Run 3                      & 64   & 0.17  & 2024-09-03  & 2024-12-24  & 42   & 0.13  & 2024-09-07  & 2024-12-17 & 4 \\
\enddata
\tablenotetext{a}{HIRES underwent a major instrument upgrade in August 2004 \citep{butler_lces_2017}. Since all HIRES data for 61~Cyg A and B were taken after this change, we do not need to split these RVs into multiple runs.}
\tablerefs{(1)~\citet{rosenthal_california_2021}; (2)~\citet{Teklu2025_morehires}; (3)~\citet{giovinazzi_neid_2025}; (4)~this work. The full RV time series are available on Zenodo (\dataset[10.5281/zenodo.22909717]{10.5281/zenodo.22909717}).}
\end{deluxetable}

We perform a similar analysis, incorporating several new RV datasets that have become available in the ten years since the work of \citet{howard_limits_2016}, which are summarized in Table \ref{tab:rvs}.
The California Legacy Survey \citep{rosenthal_california_2021} compiles data from the Hamilton and HIRES spectrographs as well as the Automated Planet Finder \citep[APF;][]{Vogt2014_apf} up through April 2020. The HIRES RV baseline can be further extended to March 2023 with an updated catalog by \citet{Teklu2025_morehires}, which utilizes a slightly different data reduction pipeline based on \citet{butler_lces_2017} and \citet{TalOr2019_hires}.
In addition, the precision RV spectrograph NEID \citep{schwab_design_2016, Halverson2016_errorbudget} has been monitoring 61~Cyg~AB since 2021, with observations predominantly coming from two long-running programs monitoring nearby RV-quiet stars for planetary signals at the m\,s$^{-1}$ level, the NEID Earth Twin Survey \citep[NETS;][]{Gupta2021_NETSoverview} and the Searching for Nearby Exoplanets Around K-dwarfs survey \citep[SNEAK;][]{lin_precision_2024}. These data become publicly accessible via the NEID Archive\footnote{\url{https://neid.ipac.caltech.edu/search.php}} once their proprietary periods expire. \citet{giovinazzi_neid_2025} used a subset of the public NEID data (corresponding to Science Run~1 as described in \S\ref{sec:rvsearch}) to characterize the long-period binary orbit of 61~Cyg AB, and also published RVs of 61~Cyg~A taken by the MINERVA telescope array \citep{Swift2015_minerva} from 2017--2019. 

\subsubsection{NEID RVs}
\label{sec:neid_rvs}

NEID is a high-resolution ($R \approx$~100,000), environmentally-stabilized \citep{Robertson2019_neidthermal} red-optical (380--930 nm) spectrograph mounted on the WIYN 3.5~m telescope\footnote{The WIYN Observatory is a joint facility of the NSF's National Optical-Infrared Astronomy Research Laboratory, Indiana University, the University of Wisconsin-Madison, Pennsylvania State University, Purdue University and Princeton University.} at Kitt Peak National Observatory. It was designed specifically for precision RVs with an instrumental error budget of $\lesssim$~30~cm\,s$^{-1}$, in order to deliver the sub-m\,s$^{-1}$ on-sky precision required for the detection of small rocky planets around nearby Sun-like stars \citep{Halverson2016_errorbudget}.

We queried the NEID Archive for publicly available data of 61~Cyg AB, and find 250 individual exposures for 61~Cyg~A on 119 nights spanning 2021 May~31 to 2024 Dec~24, and 134 exposures for 61~Cyg~B on 49 nights spanning 2021 Oct~06 to 2024 Dec~31. The Archive version of these data have been automatically processed by the NEID Data Reduction Pipeline (DRP)\footnote{\url{https://neid.ipac.caltech.edu/docs/NEID-DRP/overview.html}}, which computes RVs via the classic cross-correlation function (CCF) method \citep{Baranne1996_ccf}. In a bid to achieve the best possible RV precision, we take the DRP-processed 1D spectra and re-derive RVs via the template-matching method \citep{AngladaEscude2012_templatematching}; for previous NEID observations of late-type stars, including 61~Cyg~B itself, template-matching RVs have consistently demonstrated improved RV precision compared to CCF RVs \citep{lin_precision_2024}. To compute the template-matching RVs, we use a customized version of the SpEctrum Radial Velocity AnaLyser package \citep[\texttt{SERVAL};][]{Zechmeister2018_serval} which has been adapted for NEID as described in \citet{Stefansson2022_neidserval}, using the central 7000 pixels of NEID echelle orders 153--69 (indices 20--104, $\lambda_c$~= 401--895~nm).

\subsubsection{Planet Constraints with \texttt{RVsearch}}
\label{sec:rvsearch}

All together, these observations form RV baselines spanning 37 years for both 61 Cyg~A and B. We use \texttt{RVsearch} \citep{rosenthal_california_2021} to search for both long-term RV trends and periodic signals, and then derive detection completeness maps through injection-recovery tests. 

For the purposes of RV fitting, we split the NEID data into three subsets due to instrumental and algorithmic changes which affect the RV zeropoint, and treat each as a separate instrument (i.e., independent RV offset and jitter terms). We refer to these as Science Runs 1, 2, and 3 to match the nomenclature of the NEID DRP\footnote{\url{https://neid.ipac.caltech.edu/docs/NEID-DRP/rveras.html}} --- NEID was thermal-cycled when Kitt Peak shut down due to the Contreras Fire between Runs~1 and 2, and an upgrade to the laser frequency comb improved the wavelength solution derivation starting with Run~3. Following \citet{giovinazzi_neid_2025}, we also treat the four MINERVA component telescopes (T1--T4) as separate instruments. 
Finally, we choose to use the \citet{Teklu2025_morehires} RVs for \emph{all} of the HIRES data, since mixing the results of different pipelines would introduce an additional RV offset into the HIRES dataset. This reduction has slightly worse RV precision for 61~Cyg~A\footnote{Median error of 1.26 vs.\ 0.96~m\,s$^{-1}$ for 61~Cyg~A, while the median error for 61~Cyg~B actually improves, from 1.07 to 0.89~m\,s$^{-1}$.} compared to \citet{rosenthal_california_2021}, but we judged this to be outweighed by the advantage of a longer unbroken RV series which also overlaps in time with NEID Science Run~1, anchoring the RV offset between the two instruments which would otherwise be completely unconstrained.

\begin{deluxetable}{llll}
    \label{tab:priors}
    \tabletypesize{\footnotesize}
    \centering
    \tablecaption{Orbit Simulation Parameters and Adopted Priors}
    \tablehead{\colhead{Parameter} & \colhead{Description} & \colhead{Unit} & \colhead{Prior}} 
        \startdata
        \sidehead{Stellar Orbit}
        $M_A$ & 61~Cyg~A Mass & $M_\odot$ & $\mathcal{N}$(0.677, 0.005) \\
        $M_B$ & 61~Cyg~B Mass & $M_\odot$ & $\mathcal{N}$(0.629, 0.009) \\
        $\varpi$ & Parallax & mas & $\mathcal{N}$(286.019, 0.069) \\
        $a_{AB}$ & Semi-major Axis & au & $\mathcal{N}$(86.76, 0.16) \\
        $e_{AB}$ & Eccentricity & \nodata & $\mathcal{N}$(0.4424, 0.0057) \\
        $i_{AB}$ & Sky Inclination & deg & $\mathcal{N}$(52.99, 0.12) \\
        $\omega_{AB}$ & Argument of Periastron & deg & $\mathcal{N}$(150.61, 0.68) \\
        $\Omega_{AB}$ & Longitude of Ascending Node & deg & $\mathcal{N}$(355.73, 0.30) \\
        $t_p$ & Time of Periastron Passage & JD & $\mathcal{N}$(2596496, 1255) \\
        \sidehead{Companion Orbit}
        $a$ & Semi-major Axis & au & $\mathcal{U}$(1, 50) \\
        $m_p$ & Companion Mass & $M_{\rm Jup}$ & $\mathcal{N}$($\mathcal{F}(a)$, $\mathcal{F}_{\rm err}(a)$)\tablenotemark{a} \\
        $e$ & Eccentricity & \nodata & Population-informed (see \S\ref{sec:params}) \\
        $i$ & Sky Inclination & \nodata & Cosine \\
        $\omega$ & Argument of Periastron & rad & $\mathcal{U}$(0, 2$\pi$) \\
        $\Omega$ & Longitude of Ascending Node & rad & $\mathcal{U}$(0, 2$\pi$) \\
        $\tau$ & Time of Periastron Passage$^{15}$ & \nodata & $\mathcal{U}$(0, 1)
        \enddata
        \tablenotetext{a}{$\mathcal{F}$ is the function that defines the companion mass $m_p$ as a function of semi-major axis $a$ compatible with the velocity anomaly. $\mathcal{F}_{\rm err}$ is the function that defines the uncertainty in companion mass $m_p$ as a function of semi-major axis $a$.}
\end{deluxetable}

As expected, both stars show long-term quadratic RV trends due to the orbit of the other (Figure~\ref{fig:RVs}), and 61~Cyg~B also exhibits a periodic RV signal at $P$~= 49.03~d with a semi-amplitude of $K \approx$~2.2~m\,s$^{-1}$ that clears a false alarm probability (FAP) threshold of 0.1\% (Figure~\ref{fig:periodogram}; the \texttt{RVsearch} FAP is calculated empirically, and is further detailed in \citealt{howard_limits_2016}). Both \citet{rosenthal_california_2021} and \citet{lin_precision_2024} had detected this signal previously, and both associated it with the $\sim$48~d stellar rotation period derived from Ca~II~HK \citep{Vaughan1981_prot}. We also note much weaker periodogram peaks ($\sim$1\% FAP) in the neighborhood of 2350~d (6.4~yr) for both 61~Cyg A and B (Figure~\ref{fig:periodogram}), which do not line up with their established long-term activity cycles of 7.3 and 11.7 years, respectively \citep{Baliunas1995_activity, olspert_estimating_2018}. Given the low significance of these potential periodicities \citep[see e.g.,][]{burt_precise_2026}, we do not investigate them further in this work.

We then run injection-recovery tests spanning $P =$ 10--10$^5$~d and $K =$ 1--10$^3$~m\,s$^{-1}$ with a log-uniform distribution in both parameters, using 10,000 synthetic planets to construct each completeness map. \citet{howard_limits_2016} considered only circular orbits for their injection-recovery tests, reasoning that the vast majority ($>$~80\%) of RV-detected planets had eccentricities low enough that their orbits could be approximated as circular. We cannot make the same assumption here because recent work has shown that the $\gtrsim1$~au companion eccentricity distribution is mass-dependent and spans a broad range of values \citep{bowler_population-level_2020, nagpal_impact_2023, gilbert_orbital_2026}. Therefore, we generate RV completeness maps across a grid of eccentricity values ($e =$ 0--0.9 in steps of 0.1) to use in our analysis (Figure~\ref{fig:data}).

\section{Updated Constraints on Companion Properties}
\label{sec:constraints}

In this section, we combine the astrometric evidence for a companion in the 61 Cyg system together with existing imaging observations, new radial velocity measurements, the Gaia DR3 {\tt RUWE} value of each star, and dynamical stability calculations in a Monte Carlo simulation framework to obtain new constraints on the orbital and physical properties of the candidate companion, which we refer to as ``61~Cyg~b" for convenience. The calculations discussed below are performed separately for 61~Cyg~A and B assuming the companion orbits either one of the stars. 

\subsection{Orbital Parameter Sampling Procedure}
\label{sec:sampling}
We begin by uniformly sampling the companion semi-major axis (SMA, $a$) at $10^6$ points in the range 1--50~au. Next, for each sampled value of the companion's semi-major axis, we draw the companion's mass ($m$) from a normal distribution defined by the companion mass-SMA function compatible with the observation of the orbital velocity anomaly (Figure~\ref{fig:data}). The eccentricity ($e$) of each sampled orbit is assigned by sampling the population-level eccentricity distributions of giant planets and brown dwarfs. For 1--10~au orbits, we use the mass-dependent eccentricity distributions inferred by \citet[][see their Figure 2]{gilbert_orbital_2026} using the California Legacy Survey sample \citep{rosenthal_california_2021} combined together with Hipparcos-Gaia astrometry. For $>10$~au orbits, we use the giant planet ($\leq 15\;M_{\rm Jup}$) and brown dwarf ($> 15\;M_{\rm Jup}$) eccentricity distributions inferred by \citet{nagpal_impact_2023} using the \citet{bowler_population-level_2020} sample, specifically, the case that employs a truncated Gaussian hyperprior on the beta distribution parameters when performing hierarchical Bayesian modeling. The companion's sky inclination ($i$) is uniformly sampled from a cosine distribution consistent with random orbital orientation. The remaining orbital elements of the companion (argument of periastron $\omega$, longitude of ascending node $\Omega$, and time of periastron passage parameter $\tau$\footnote{$\tau$ is a dimensionless number between 0--1 which expresses the time of periastron passage as a fraction of the companion's orbital period with respect to a given reference epoch (see \url{https://orbitize.readthedocs.io/en/latest/faq/Time_Of_Periastron.html}, for example).}) are sampled from uniform distributions spanning their possible values. The stellar mass and orbital parameters are sampled from the distributions presented in \citet{giovinazzi_neid_2025}. All priors are summarized in Table~\ref{tab:priors}.

\subsection{Dynamical Stability Simulations}
\label{sec:dynamics}

The sampled orbits are evaluated for dynamical stability over a timescale of 1 Myr, where previous studies of planets in binaries showed that instabilities typically occur on the secular timescale \citep{quarles_long-term_2016,quarles_orbital_2020,quarles_main-sequence_2026} . We use the $N$-body simulation package \texttt{Rebound}[4.4.6] \citep{Rein2012} to evolve three-body systems consisting of the 61~Cyg~AB system along with the putative companion. The posterior outputs from the above analysis provides the correlated initial conditions for these stability simulations, which include the masses (of all three bodies) and the orbital elements of the companion and stellar orbits. The system is evolved using the binary orbital plane as the reference, where both orbits start at their respective pericenter (mean anomaly, ${\rm MA} = 0^\circ$). The orbits of the companion and binary begin misaligned (in both the eccentricity and inclination vectors) so that any bias from our initial mean anomaly is negligible. 

\begin{figure*}
    \centering
    \includegraphics[width=\linewidth]{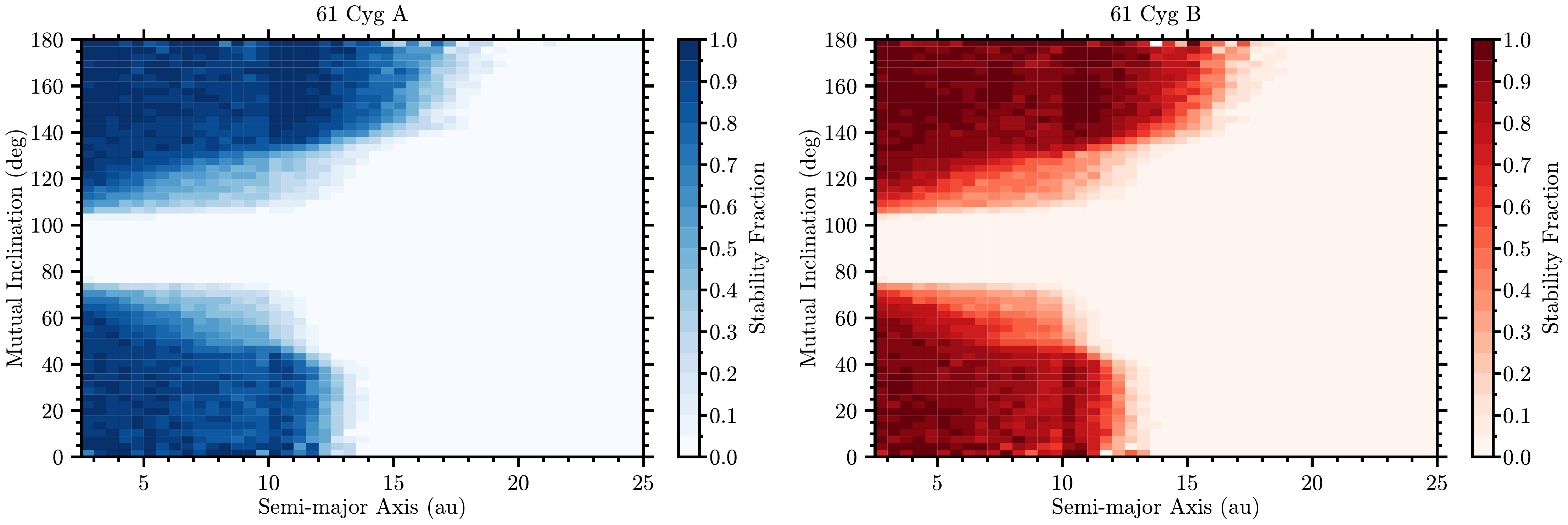}
    \caption{Two-dimensional histograms showing the stability fraction for the $10^6$ sampled orbits, which are binned in mutual inclination (relative to the binary orbital plane) vs.~semi-major axis space for 61~Cyg~A (left panel) or 61~Cyg~B (right panel) as the host star. The stability fraction is defined as the number of stable orbits divided by the total number of orbits evaluated in a given bin. High mutual inclination (near $90^\circ$) orbits are disfavored due to the von Ziepel-Lidov-Kozai mechanism \citep{vonZeipel1910,Lidov1962,Kozai1962}.  The stability fraction varies depending on wheter prograde or retrograde orbits are assumed, where $\gtrsim 12$~au or $\gtrsim16$~au are disfavored, respectively.}
    \label{fig:stability-region}
\end{figure*}

\begin{figure*}
    \centering
    \includegraphics[width=0.95\linewidth]{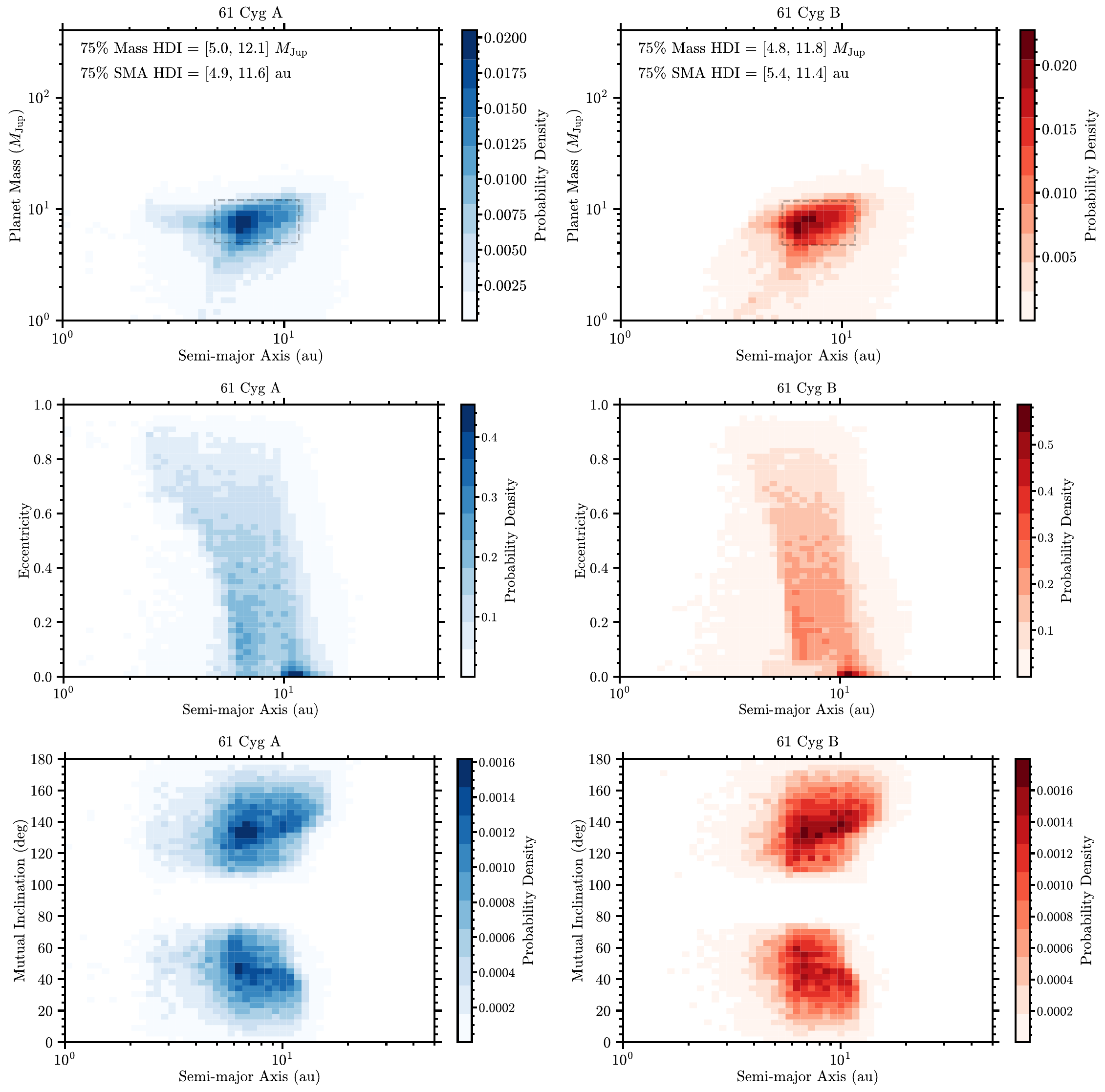}
    \caption{Posteriors for 61~Cyg~b's mass, eccentricity, and mutual inclinaton with respect to the binary orbital planet vs.~semi-major axis compatible with the velocity anomaly, Gaia DR3 {\tt RUWE}, RV and imaging non-detections, and dynamical stability. The 75\% highest density intervals (HDIs) are quoted for mass and semi-major axis and represented visually as a dashed box. The data is consistent with a super-Jupiter companion on solar system orbital scales.}
    \label{fig:final-constraints}
\end{figure*}

\begin{figure*}
    \centering
    \includegraphics[width=\linewidth]{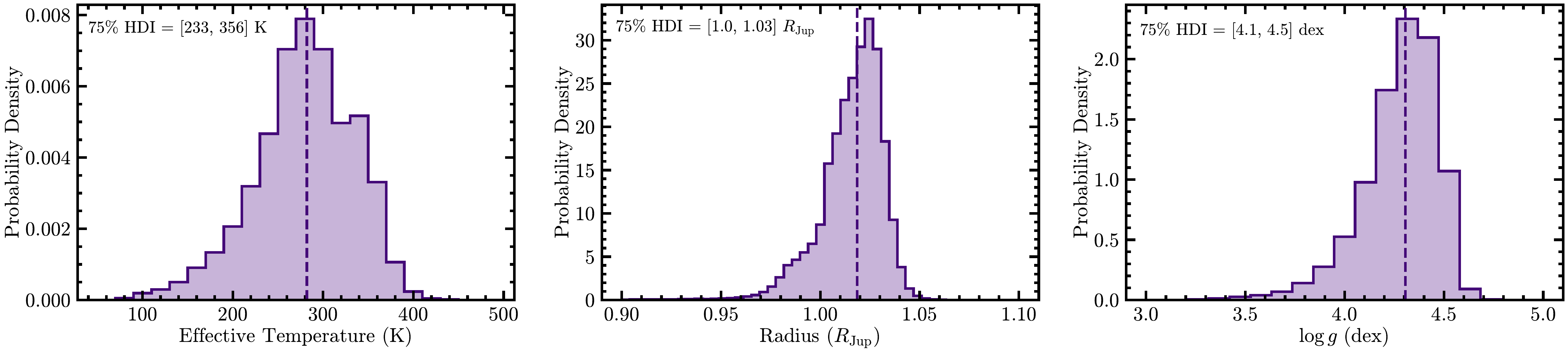}
    \caption{Posteriors for the effective temperature (left), radius (center), and surface gravity (right) of 61~Cyg~b derived using planet mass and age samples combined with the [M/H]~$=+0.5$~dex, C/O~$=1.0\times$ solar, disequilibrium chemistry Sonora Flame Skimmer evolutionary model (nearly identical posteriors for planet orbiting star A or star B). The dashed line shows the median parameter value. The 75\% highest density interval (HDI) is noted in the top left corner of each panel.}
    \label{fig:evo-prop}
\end{figure*}

Our simulations for these three-body systems use the adaptive \texttt{ias15} integrator \citep{Rein2015}, where we set a timestep floor at $10^{-4}$ times the companion's initial orbital period.  We deem an initial condition as unstable when a simulation reaches the following conditions: (1) companion collides with the host star; (2) companion eccentricity exceeds $0.95$; or (3) companion's distance exceeds the distance to the Lagrange point $L_1$. The simulation uses the ``direct" collision detection method within \texttt{Rebound} using the radii: $0.667\ R_\odot$ for 61~Cyg~A, $0.594\ R_\odot$ for 61~Cyg~B, and $1.1\ R_{\rm Jup}$ for the candidate companion. We limit the companion eccentricity because tidal effects could become important for such orbits over long timescales, which we are not including in our numerical model, as well as for numerical convenience. Companion orbits that go beyond the middle Lagrange points $L_1$ are chaotic and ultimately unstable in most instances \citep{wiegert_stability_1997,quarles_orbital_2020}. Only very special conditions \citep[e.g., net zero angular momentum;][]{Suvakov2013} allow for stable orbits that cross $L_1$. 

The results of the stability simulations are shown in Figure~\ref{fig:stability-region}, where it is evident that orbits near polar configurations ($\approx$80$^\circ$--100$^\circ$) are entirely unstable.  Prograde orbits with $40^\circ \lesssim i_{\rm mut} \lesssim 75^\circ$ allow for a stable fraction of $0.5$ for a planetary semi-major axis between ${\sim}5$--$10$~au, while the complementary retrograde orbits extend farther for the same stability fraction to ${\sim}5$--$12$~au.  For nearly coplanar, prograde planetary orbits ($0\leq i_{\rm mut}\lesssim40^\circ$) the stability fraction drops significantly beyond ${\sim}12$~au, while the retrograde counterparts ($140^\circ \lesssim i_{\rm mut} \leq 180^\circ$) extend to ${\sim}16$~au with a relatively high stability fraction.  Most bins do not achieve stability fraction exactly equal to $1$ due to the wide range of planet eccentricity present in the orbit sample. 

\subsection{61 Cyg b's Mass and Orbit}
\label{sec:params}

The sampled orbits are first triaged using constraints from the Gaia DR3 {\tt RUWE} of each star (Figure~\ref{fig:data}). We reject all orbits (indexed by $j$) where $P({\tt RUWE}_{\rm sim}\leq{\tt RUWE}_{\rm obs}\mid m_{p_j}$,\;$a_j) \leq 5\%$. Next, we apply the eccentricity-dependent RV completeness maps $f(m\sin i,\;a,\;e)$ generated in \S\ref{sec:RV} (Figure~\ref{fig:data}). The eccentricity of each sampled orbit ($e_j$) is rounded to the nearest tenth to determine which completeness map is applied. We eliminate all orbits where the planet would have been recovered by RV observations with $>$95\% completeness, or equivalently, $f(m_j\sin i_j,\;a_j,\;e_j) > 0.95$. Note that no orbit with $e_j \geq 0.85$ is eliminated, as there is no region where $>95\%$ RV completeness is achieved for these eccentricity values. For the accepted orbits, we compute the companion's projected separation on the dates of Clio $Lp$ imaging. We then reject orbits where the companion would have been recovered by Clio imaging according to the mass sensitivity curves in \S\ref{sec:imaging}. Finally, all unstable orbits are removed following the results of the $N$-body dynamical stability simulations in \S\ref{sec:dynamics}.

The key properties of the surviving orbits (available on Zenodo\dataset[10.5281/zenodo.22909717]{10.5281/zenodo.22909717}), consistent with the tangential velocity anomaly, Gaia DR3 {\tt RUWE}, non-detection in RV and imaging observations, and stability in the system are shown in Figure~\ref{fig:final-constraints}. The mass and semi-major axis of the companion are $m_p = 8.5\pm3.0\;M_{\rm Jup}$, $a = 8.2\pm3.0$~au if orbiting 61~Cyg~A and $m_p = 8.2\pm3.0\;M_{\rm Jup}$, $a = 8.5\pm2.7$~au if orbiting 61~Cyg~B. The 75\% highest density interval for the posterior is noted in Figure~\ref{fig:final-constraints}. The eccentricity distribution broadly reflects the population-level distribution \citep{nagpal_impact_2023, gilbert_orbital_2026} that was applied as a prior, favoring values $e \lesssim 0.5$. The expected eccentricity of 61~Cyg~b does, however, increase for smaller semi-major axis values (Figure~\ref{fig:final-constraints}) because of the poor completeness of the RV data to high eccentricity orbits (Figure~\ref{fig:data}). The mutual inclination ($i_{\rm mut}$) distribution is bimodal, shaped by the dynamically stable regions identified in Figure~\ref{fig:stability-region}, and allows for both prograde ($i_{\rm mut} = 41^\circ \pm 17^\circ$) and retrograde ($i_{\rm mut} = 141^\circ \pm 17^\circ$) solutions (similar for both stars).

\subsection{61 Cyg b's Effective Temperature and Radius}
\label{sec:teff-rad}
Given the mass and age of 61~Cyg~b, we derive the planet's effective temperature and radius using the Sonora Flame Skimmer atmospheric and evolutionary models \citep{mang_sonora_2026}. These extend the cloud-free Sonora Elf Owl grid \citep{mukherjee_sonora_2024, wogan_sonora_2025} to colder effective temperatures, lower surface gravities, and a broader range of metallicities. The models include both equilibrium and disequilibrium chemistry and incorporate rainout chemistry for H$_2$O, CH$_4$, and NH$_3$—even in cloud-free atmospheres—similar to the treatment in Sonora Bobcat \citep{marley_sonora_2021}. They also address the underestimation of CO$_2$ found in the Sonora Elf Owl models \citep{mukherjee_sonora_2024}, which has since been revised in \citet{wogan_sonora_2025}. The evolutionary models use the clear atmospheric models as surface boundary conditions and span masses from 15~$M_\oplus$ to 0.1~$M_\odot$, with metallicities of [M/H] = {$-$1.0, $-$0.5, $+$0.0, $+$0.5, $+$1.0, $+$1.5, $+$2.0} dex. For objects with masses below 3~$M_{\rm Jup}$, the models include a 15~$M_\oplus$ core in the interior. These evolutionary models adopt the hydrogen and helium equations of state from \citet{chabrier_new_2019} and \citet{chabrier_new_2021}, and use the water equation of state from \citet{mazevet_ab_2019} to account for all metals. The effects of these choices are discussed in \citet{chachan_revising_2025}. For the calculations presented here, we use the [M/H]~$=+0.5$~dex, C/O~$=1.0\times$ solar, disequilibrium chemistry Sonora Flame Skimmer evolutionary model, which is representative of a giant planet scenario. We use the mass samples of 61~Cyg~b derived previously (Figure~\ref{fig:final-constraints}, nearly identical results with star A and star B) and assume a normal distribution for age ($6.0 \pm 1.0$~Gyr). The evolutionary models are linearly interpolated at each (mass, age) sample to derive the effective temperature ($T_{\rm eff}$), radius ($R$), and surface gravity ($\log g$, cgs units). The posterior distributions are well constrained and we find a $T_{\rm eff} = 282 \pm 55$~K, $R = 1.02 \pm 0.02\;R_{\rm Jup}$, and $\log g = 4.3 \pm 0.2$~dex (Figure~\ref{fig:evo-prop}) for 61~Cyg~b. Note that the quoted uncertainties are only statistical in nature. Systematic uncertainties in the model are not known, but are likely to be small given that the fundamental parameters predicted by evolution models for a comparably cold and old planet, $\epsilon$~Ind~Ab, are in agreement with its empirical measurements \citep{sanghi_worlds_2026-1}. 

\section{Stability of Terrestrial Planets in the Habitable Zone}
\label{sec:stable-earths}

61~Cyg~AB have been identified as prime targets to search for habitable zone exo-Earths with future missions such as SHERA, HWO, and LIFE. We perform $N$-body simulations with the \texttt{Rebound}[4.4.6] \citep{Rein2012} package to assess how the inclusion of the candidate giant planet affects terrestrial planets in the \textit{empirical} habitable zone of the host star. This is done by evolving a system of 100 massless test particles in the empirical habitable zone along with the candidate giant planet and binary companion for their respective orbital parameters. These parameters are drawn from the final set of triaged orbits determined in Section~\ref{sec:params}. As a result, it is a sample that spans a wide range of correlated parameters in the candidate giant planet's eccentricity, semi-major axis, and mutual inclination. The simulation results are expected to be similar for both stars since the candidate giant planet's properties (Figure~\ref{fig:final-constraints}) and the stellar spectral type are similar. Here, we perform the calculations assuming the candidate giant planet orbits 61~Cyg~A. 

The empirical habitable zone uses the flux received by Venus ($1.91\ S_\oplus$) and Mars ($0.32\ S_\oplus$) as the inner and outer limits \citep{Kasting1993}.  For 61 Cyg A, we find that the inner and outer distances are $0.28$~au and $0.68$~au, respectively, using the zeroth-order formula that depends on the host star's luminosity $L/L_\odot$ and the effective flux received $S_{\rm eff}$ at that distance.  The simulations are evolved for the same timescale of 1 Myr used for the dynamical stability simulations, but use the \texttt{TRACE} integrator instead of \texttt{ias15}.  

The orbital periods of the test particles are much shorter, where the \texttt{TRACE} integrator allows us to evaluate their initially Keplerian orbits more efficiently at the beginning of the simulation.  The \texttt{WHFast} part of the \texttt{TRACE} algorithm uses a fixed timestep of $0.025\times$ the orbital period at the inner edge of the habitable zone (i.e., $2.5\%$ of the shortest orbital period).  A test particle is removed from the simulation if: (1) it collides with either star or the candidate planet, (2) it has an eccentricity that exceeds $0.6$ for at least 100 yr, or evolve to a distance beyond the $L_1$ Lagrange point.  These removal criteria are effectively the same as those we used for our stability simulations, but the second criterion is a more targeted application of a limiting eccentricity based on the habitability concerns \citep{bolmont_habitability_2016}.  Due to our removal criteria, using the \texttt{ias15} is slower and unnecessary because the test particles are removed before they evolve to very high eccentricity.

There are 4 particles per ring and 25 rings that are uniformly spaced throughout the habitable zone.  The test particles begin from an initially circular orbit that is coplanar with the binary orbit, while the mean anomaly of the test particles are offset by half a phase relative to the neighboring ring.   This offset pattern produces a more uniform area of test particles throughout the habitable zone.  The resulting metric at the end of the simulation is the fraction of the original habitable zone particles remaining.  If there are $0\%$ remaining, then the candidate giant planet is likely efficient at disrupting planet formation from occurring within the habitable zone \citep{Marzari2000,Thebault2006,raymond_search_2006,thebault_planet_2008}. In contrast, if there is a large percentage remaining, then the habitable zone can be perturbed but can otherwise still potentially host terrestrial planets.

\begin{figure}
    \centering
    \includegraphics[width=\linewidth]{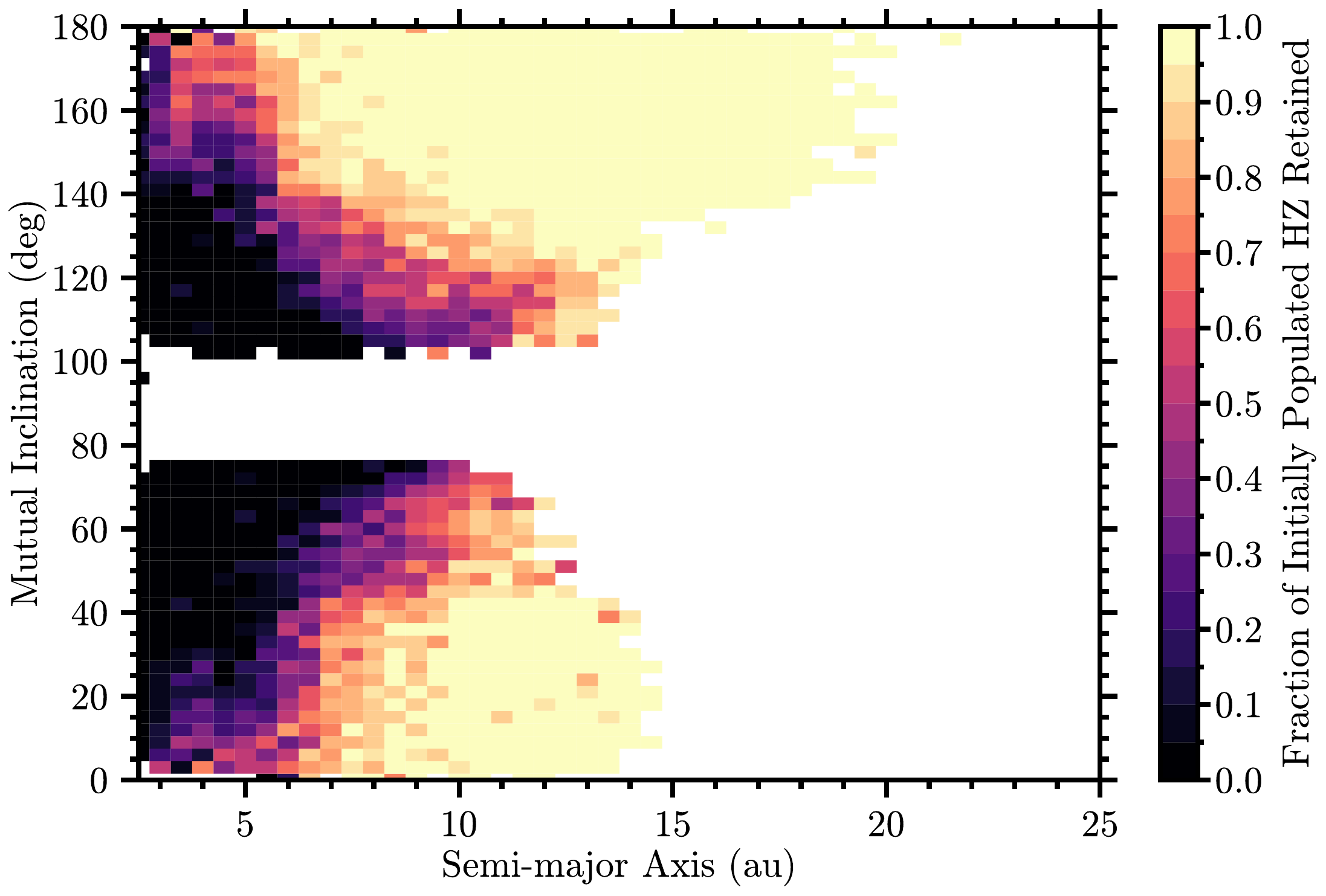}
    \caption{Fraction of HZ test particles that survive the 1~Myr dynamical simulations in the presence of the giant planet candidate (assumed to be orbiting 61~Cyg~A here) as a function of the giant planet candidate's semi-major axis and mutual inclination with respect to the stellar binary or, equivalently, the HZ test particles (by simulation design).}
    \label{fig:HZ}
\end{figure}

The results of the $N$-body simulations for 61~Cyg~A are shown in Figure~\ref{fig:HZ} (similar for 61~Cyg~B, as noted earlier), where we bin the results in the mutual inclination and semi-major axis of the giant planet candidate (similar to Fig. \ref{fig:stability-region}). Giant planet candidate orbits with semi-major axis $\lesssim5$~au strongly disfavor the stability of test particles in the HZ, across all mutual inclinations, as the planetary orbits are also highly eccentric (Figure~\ref{fig:final-constraints}), and thus, have a periastron passage close to the empirical HZ. Larger semi-major axis orbits ($\gtrsim5$~au, up to $\sim$10~au) are dynamically unfavorable for misaligned configurations ($ 40^\circ \lesssim i_{\rm mut} \lesssim 140^\circ$) as they lead to the disruption of HZ test particles by driving von Ziepel-Lidov-Kozai oscillations \citep{vonZeipel1910,Lidov1962,Kozai1962,2016ARA&A..54..441N}.

\section{Conclusions}
\label{sec:concl}

In this work, we presented new constraints on the properties of a candidate companion, identified using Hipparcos-Gaia DR3 astrometry, in the 61~Cygni~AB binary system by combining absolute astrometry, RV measurements, archival direct imaging data, and dynamical stability calculations for the two stars. The key takeaways are summarized below:

\begin{enumerate}
    \item \textbf{New NEID RV data extends the RV monitoring baseline of 61~Cyg~AB to 37~years.} We performed a comprehensive analysis of RV constraints on possible companions in the system using \texttt{RVSearch}, with the addition of new publicly available data from APF, MINERVA, and NEID, and an updated HIRES RV catalog from \citet{Teklu2025_morehires} which extends the HIRES coverage through October 2022. No signals attributable to a third companion were robustly detected. We performed injection-recovery tests and generated RV detection completeness maps as a function of companion eccentricity to help constrain the Hipparcos-Gaia candidate's properties in our joint analysis.
    
    \item \textbf{Regions of dynamical stability for the candidate companion exist in the binary system.} Using $N$-body simulations performed over a 1~Myr timescale, we found that only near polar ($80^\circ \lesssim i_{\rm mut} \lesssim 100^\circ$) or large semi-major axis ($\gtrsim12$~au for the prograde and $\gtrsim$16~au for the retrograde cases) orbits are dynamically excluded for the candidate in the 61~Cyg~AB system. The remaining parameter space allowed stable configurations for the candidate companion around either star.
    
    \item \textbf{A cold super-Jupiter candidate in a solar system scale orbit is consistent with all data in hand.} We simulated orbits randomly oriented in space, consistent with the tangential velocity anomaly signal, and removed those inconsistent with the Gaia DR3 {\tt RUWE}, RV and MMT/Clio $Lp$ direct imaging limits, and dynamical stability limits for the two stars. The final triaged orbit sample constrained the candidate companion's ($m_p$, $a$) to $(8.5\pm3.0\,M_{\rm Jup}, 8.2\pm3.0\,\rm{au})$ if orbiting 61~Cyg~A, and $(8.2\pm3.0\,M_{\rm Jup}, 8.5\pm2.7\,\rm{au})$ if orbiting 61~Cyg~B. For an age of $(6.0 \pm 1.0)$~Gyr, we found an effective temperature of $(282 \pm 55)$~K and radius of $(1.02 \pm 0.02)\;R_{\rm Jup}$ for the candidate using the Sonora Flame Skimmer evolutionary models.

    \item \textbf{The giant planet candidate may dynamically disrupt terrestrial planets in the HZ of 61~Cyg~AB.} We populated the empirical HZs of 61~Cyg~A and B with massless test particles and evolved them together with the giant planet candidate and stellar binary (accounting for the new orbital constraints) over a 1~Myr timescale using $N$-body simulations. We found that if the giant planet candidate's semi-major axis is $\lesssim5$~au, then, irrespective of the mutual inclination between the test particles and the giant planet candidate, the HZ is heavily disrupted. Between $\approx$5~au and 10~au, the giant planet candidate can also destabilize the HZ for misaligned orbits ($40^\circ \lesssim i_{\rm mut} \lesssim 140^\circ$) due to von Ziepel-Lidov-Kozai oscillations. 
\end{enumerate}

The upcoming release of Gaia DR4 is expected to improve the precision of 61~Cyg~AB's proper motion by a factor of three to four\footnote{Based on a \href{https://great.ast.cam.ac.uk/Greatwiki/GreatMeet-PM19?action=AttachFile&do=view&target=EAS2026-S7-Brown.pdf}{presentation} by A.~Brown at the Great Plenary Meeting: PM19 / EAS Science Symposium S7.} and include an astrometric timeseries for the two stars. A follow-up study to this paper will incorporate the above data into our analysis framework with the goal of improving the significance of the astrometric signal and parameter constraints for the candidate companion. Additional RV and/or direct imaging observations will be required to confirm or rule out the candidate given that the expected orbital period (more than 10~years) is longer than Gaia DR4's baseline (5.5~years). Such efforts are critical as the possible presence of a giant planet has important implications for the long-term survival of habitable zone terrestrial planets in the 61~Cyg~AB system, and thus, Earth-analog detection and characterization campaigns with future missions such as SHERA, HWO, and LIFE. 

\begin{acknowledgments}
A.S thanks Lynne Hillenbrand for helpful discussion on the age of 61~Cyg~AB, Brendan Bowler for guidance on population-level constraints on the giant planet and brown dwarf eccentricity distribution, Greg Gilbert for sharing the eccentricity distributions inferred from the CLS sample, and his Ph.D.~candidacy committee---Andrew Howard, Dimitri Mawet, Heather Knutson, Kareem El-Badry, Charles Beichman, and Phil Hopkins---for motivating this paper.

A.S acknowledges support from the National Science Foundation Graduate Research Fellowship under Grant No.~2139433. J.W.X. acknowledges support from the Heising-Simons Foundation 51 Pegasi b Fellowship (grant \#2025-5887). MW was supported by the Science and Technology Facilities Council grant UKRI1198. L.M. acknowledges funding support from the DAE through the NISER project RNI 4011. This work benefited from the 2026 Exoplanet Summer Program in the Other Worlds Laboratory (OWL) at the University of California, Santa Cruz, a program partially supported by funding from NASA. Computational resources were provided by the Texas A\&M High Performance Research Computing (HPRC) facility through the Launch cluster, supported by the NSF under grant No. 2232895. This work obtained access to the Launch cluster at Texas A\&M HPRC through allocation PHY240337 from the Advanced Cyberinfrastructure Coordination Ecosystem: Services \& Support (ACCESS) program, which is supported by NSF grants 2138259, 2138286, 2138307, 2137603, and 2138296.

Based in part on observations at Kitt Peak National Observatory, NSF's NOIRLab (Prop. IDs 2021A-0390, 2021B-0450, 2022A-494327, 2023B-121504, and 2024A-416259, PI: J.~Wright; 2021A-2015, 2021B-2015, 2022A-2015, 2022B-2015, 2023A-2015, 2023B-2015, and 2024B-418185, PI: S.~Mahadevan; 2021B-0225, 2021B-0439, 2022A-174847, 2022A-923895, 2023A-621448, 2023B-936288, 2023B-981173, 2024A-820750, 2024B-422321, and 2024B-543619, PI: A.~Lin), managed by the Association of Universities for Research in Astronomy (AURA) under a cooperative agreement with the National Science Foundation. The authors are honored to be permitted to conduct astronomical research on Iolkam Du\'ag (Kitt Peak), a mountain with particular significance to the Tohono O'odham. Data presented herein were obtained at the WIYN Observatory from telescope time allocated to NN-EXPLORE through the scientific partnership of the National Aeronautics and Space Administration, the National Science Foundation, and the US National Optical-Infrared Astronomy Research Laboratory.
We thank the NEID Queue Observers and WIYN Observing Associates for their skillful execution of our NEID observations.
\end{acknowledgments}

\begin{contribution}
A.~Sanghi conceived the idea for this project, assembled the team and coordinated the overall analysis, incorporated the imaging constraints, conducted the orbital Monte Carlo simulations, determined 61~Cyg~b's properties, and led the writing and submission of the manuscript. B.~Quarles performed the $N$-body dynamical stability simulations and authored Section~\ref{sec:dynamics} and Section~\ref{sec:stable-earths}. A.~S.~J.~Lin led the radial velocity data reduction and analysis and authored Section~\ref{sec:RV}. C.~Shariat conducted the {\tt gaiamock RUWE} simulations and authored Section~\ref{sec:RUWE}. P.~Kervella derived the companion mass-semimajor axis curve implied by the tangential velocity anomaly. J.~Mang generated the Sonora Flame Skimmer atmospheric models. D.~Mawet provided early guidance and feedback on this work. M.~R.~Giovinazzi and A.~F.~Gupta contributed to acquiring the NEID radial velocity data used in this work. All co-authors provided feedback and discussions on the ideas in the manuscript.
\end{contribution}

\facilities{Hipparcos, Gaia, MMT (Clio), Lick (Hamilton, APF), Keck (HIRES), MINERVA, WIYN (NEID)}

\software{\texttt{astropy
} \citep{astropy_collaboration_astropy_2013, astropy_collaboration_astropy_2018, astropy_collaboration_astropy_2022}, \texttt{matplotlib
} \citep{hunter_matplotlib_2007}, \texttt{numpy
} \citep{harris_array_2020}, \texttt{pandas
} \citep{mckinney_data_2010, team_pandas-devpandas_2025}, \texttt{python
} \citep{van_rossum_python_2009}, \texttt{scipy
} \citep{virtanen_scipy_2020, gommers_scipyscipy_2023}, \texttt{astroquery
} \citep{ginsburg_astroquery_2019, ginsburg_astropyastroquery_2024}, \texttt{scikit-image
} \citep{van_der_walt_scikit-image_2014}, and
\texttt{Rebound} \citep{Rein2012}. Anthropic Claude Opus 5 was used to assist code development and figure preparation following the AAS guidelines as stated at \url{https://journals.aas.org/author-llm-guidelines}. The authors verified all results and accept responsibility for the intellectual integrity of the work presented.}

\appendix
\restartappendixnumbering

\section{61 Cyg AB RV Fits} 
\label{app:rv}
\begin{figure*}[!h]
    \centering
    \includegraphics[width=\linewidth]{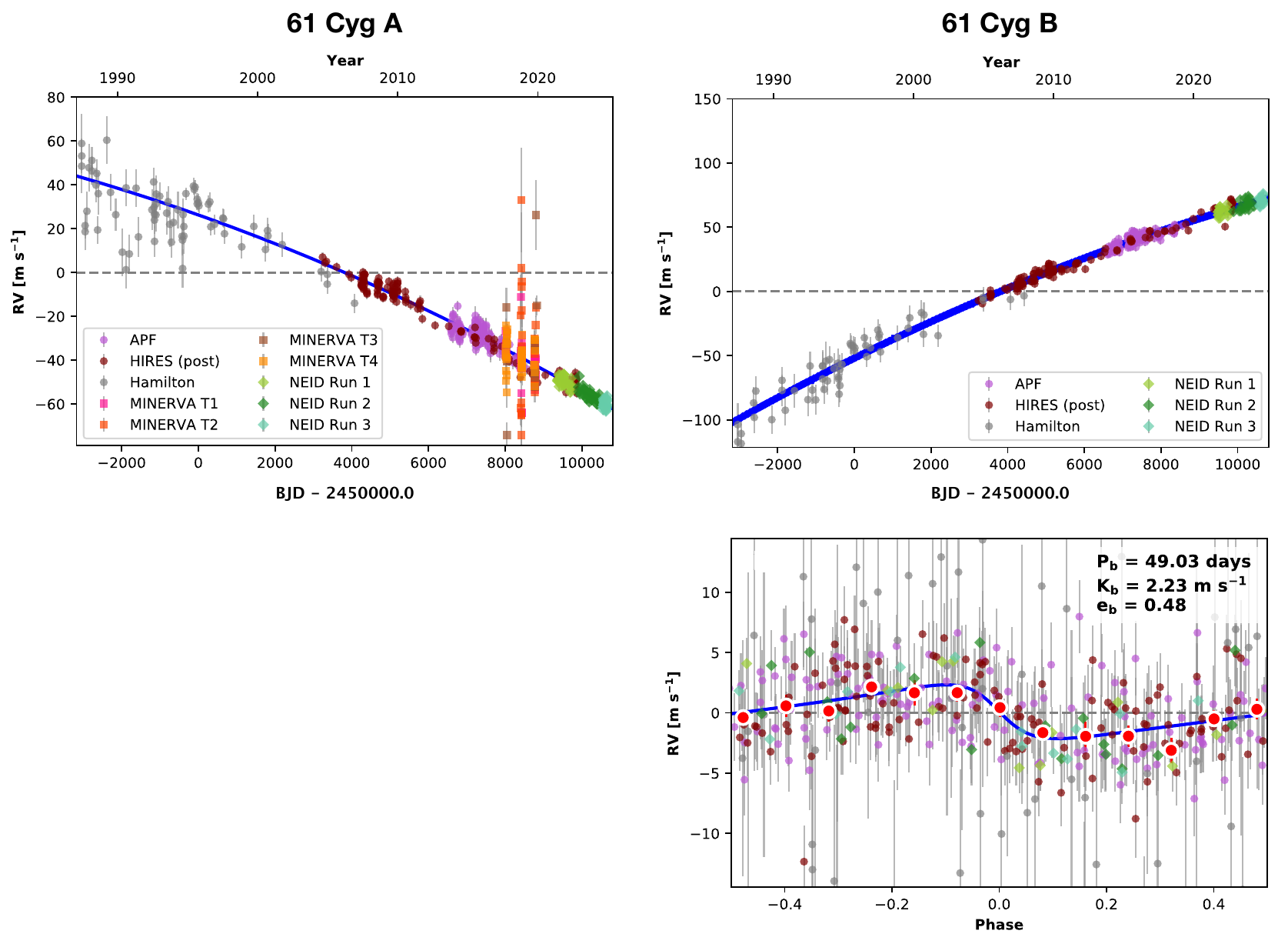}
    \caption{RVs of 61~Cyg A and B, with the blue line indicating the best-fit RV model. The binary orbit of the two stars produces quadratic long-term trends in both sets of RVs, and 61~Cyg~B also has a $\sim$2~m\,s$^{-1}$ signal at 49~d which has been attributed to the stellar rotation period (the large red points in this panel are binned by phase).}
    \label{fig:RVs}
\end{figure*}

\begin{figure*}[!h]
    \centering
    \includegraphics[width=\linewidth]{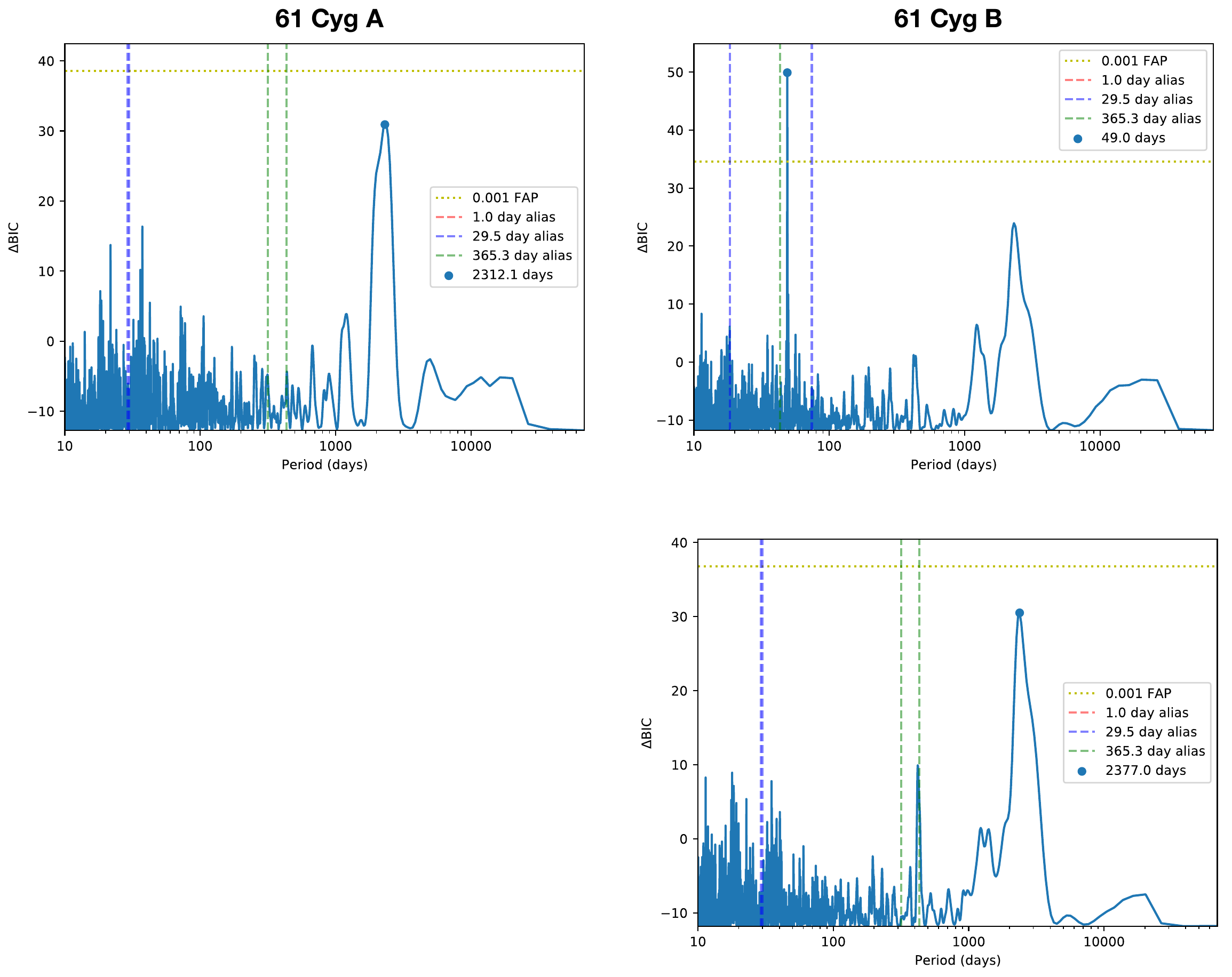}
    \caption{\texttt{RVsearch} periodograms for 61~Cyg A and B. The detected 49~d RV signal of 61~Cyg~B has an empirical FAP $<$~0.1\%, and we see subsequent weaker periodicities around 2350~d for both stars which do not clear this threshold but may merit further investigation in future work. Dashed vertical lines mark where aliasing of the primary peak with daily, lunar, and annual observing cycles may manifest as additional power in the periodogram.}
    \label{fig:periodogram}
\end{figure*}

We present the the fits to the complete RV datasets of 61~Cyg~AB and results of the \texttt{RVsearch} periodogram search in Figures~\ref{fig:RVs} and \ref{fig:periodogram}, respectively.

\newpage
\bibliography{references,references_Quarles,references_ASJL}{}
\bibliographystyle{aasjournalv7}

\end{document}